\documentstyle[preprint,aps,graphicx,eqsecnum]{revtex}

\begin{document}

\title{Universal Texture of 
Quark and Lepton  Mass Matrices }

\author{Yoshio KOIDE}
\address{
Department of Physics, 
University of Shizuoka, 52-1 Yada, Shizuoka, 422-8526 Japan \\
E-mail: koide@u-shizuoka-ken.ac.jp}
\author{Hiroyuki NISHIURA}
\address{
Department of General Education, 
Junior College of Osaka Institute of Technology, 
Asahi-ku, Osaka,535-8585 Japan \\
E-mail: nishiura@jc.oit.ac.jp}
\author{Koichi MATSUDA, Tatsuru KIKUCHI, and Takeshi FUKUYAMA}
\address{
Department of Physics, 
Ritsumeikan University, Kusatsu, Shiga, 525-8577 Japan \\
E-mail: sph30101@se.ritsumei.ac.jp, rp009979@se.ritsumei.ac.jp, 
and fukuyama@se.ritusmei.ac.jp}

%\date{\today}

\maketitle

\begin{abstract}
Against the conventional picture that the mass matrix forms in the
quark sectors will take somewhat different structures from those in the 
lepton sectors,  a possibility that all the mass 
matrices of quarks and leptons have the same form as in the 
neutrinos is investigated.
For the lepton sectors, the model leads to a nearly bimaximal mixing 
with the prediction $|U_{e3}|^2=m_e/2m_\mu=0.0024$ and 
$\tan^2\theta_{sol} \simeq m_{\nu 1}/m_{\nu 2}$, and so on. 
For the quark sectors, it can lead to reasonable values of 
the CKM mixing matrix and masses:
$|V_{us}|\simeq \sqrt{m_d/m_s}$, $|V_{ub}|\simeq |V_{cb}|\sqrt{m_u/m_c}$,
$|V_{td}| \simeq |V_{cb}|\cdot |V_{us}|$, and so on.
\end{abstract}

%%%%%%%%%%%%%%%%%%%%%%%%%%%%%%%%%%%%%%%%%%%%%%%%%%%%%%%%%%%%%%%%%%%%%%
%\begin{multicols}{2}

%\narrowtext
\section{Model}

Recent neutrino oscillation experiments\cite{skamioka,sno,chooz,kamland} 
have highly
suggested a nearly bimaximal mixing $(\sin^2 2\theta_{12}\sim 1$, 
$\sin^2 2\theta_{23}\simeq 1)$ together with a small ratio 
$R\equiv \Delta m^2_{12}/\Delta m^2_{23} \sim 10^{-2}$.
On the other hand,  we know that
the observed quark mixing matrix $V_{CKM}$
is characterized by small mixing angles.
Thus, the mixing matrices of quarks and leptons
are very different from each other.
Therefore, usually, the following picture is
accepted:
the mass matrix forms
in the quark sectors will take somewhat different
structures from those in the lepton sectors. 

%%%%%%%%%%%%%%%%%%%%%%%%%%%%%%%%%%%%

Against such a conventional picture,
we investigate a possibility that
 all the mass matrices
of quarks and leptons have the same forms as in the
neutrino sector:
\begin{equation}
M_f = P_{Lf}^\dagger \widehat{M}_f 
P_{Rf} \ ,  
%\eqno(1) 
\end{equation}
\begin{equation}
\widehat{M}_f=
\left(
\begin{array}{lll}
\ 0 & \ a_f & \ a_f \\
\ a_f & \ b_f & \ c_f \\
\ a_f & \ c_f & \ b_f \\
\end{array}
\right) \ \, \left(f=u,d,\nu,e\right),
%\eqno(2)
\end{equation}
where $P_{f} = {\rm diag}( e^{i\delta_1^f}, e^{i\delta_2^f},
e^{i\delta_3^f})$, and  
the mass matrix $\widehat{M}_f$ is invariant
under a permutation symmetry between second and third
generations.
The mass matrix parameters $a_f$, $b_f$, and $c_f$ 
can be expressed 
in terms of the mass eigenvalues as 
$a_f = \sqrt{ {m_{f2} m_{f1}}/{2} }$, 
$b_f =( m_{f3}/2) \left[ 1+({m_{f2}-m_{f1}})/m_{f3} \right]$, 
and $c_f =- (m_{f3}/2) \left[ 1-({m_{f2}-m_{f1}})/m_{f3} \right]$.

The mass matrix form (2) was suggested from the neutrino mass
matrix form \cite{Fukuyama}
which leads to  a nearly bimaximal mixing
\begin{equation}
U_\nu\equiv
\left(
\begin{array}{ccc}
{ c_\nu}& { s_\nu}& {0} \\
{-\frac{s_\nu}{\sqrt{2}}}&
{\frac{c_\nu}{\sqrt{2}}}&
{-\frac{1}{\sqrt{2}}} \\
{-\frac{s_\nu}{\sqrt{2}}}&
{\frac{c_\nu}{\sqrt{2}}}&
{\frac{1}{\sqrt{2}}}
\end{array}
\right) \ , \
%(3)
\end{equation}  
where $U_\nu^T {M}_\nu U_\nu= {\rm diag}(-m_{\nu 1}, m_{\nu 2}, m_{\nu 3})$
and 
\begin{equation}
c_\nu = \cos\theta_\nu =
\sqrt{\frac{m_{\mu 2}}{m_{\nu 2}+m_{\nu 1}}}\ , \ \ \ \
s_\nu = \sin\theta_\nu =
\sqrt{\frac{m_{\nu 1}}{m_{\nu 2}+m_{\nu 1}}} \ .
%(4)
\end{equation}

Note that
the matrix form (1) with (2) is almost invariant under
the renormalization group equation (RGE) effects,
so  that we can use the expression (1) with (2) 
for the predictions of the physical quantities in the low-energy 
region, as well as those at the unification scale.
The zeros in this mass matrix are constrained by a discrete 
symmetry Z$_3$ that is discussed in Ref.\cite{YK-Nishi}, 
defined at a unification scale (the scale does not always mean 
``grand unification scale"). This discrete symmetry Z$_3$ is  
broken below $\mu=M_R$, at which the right-handed neutrinos acquire
heavy Majorana masses.
Therefore, the matrix form (1) will, in general, be  changed by
the RGE effects.
Nevertheless, we can use the expression (1) with (2) for the predictions 
of the physical quantities in the low-energy region, as discussed
in Ref.~\cite{YK-Nishi}.

%%%%%%%%%   chap 2 %%%%%%%%%%%%%%%
\section{Quark mixing matrix}

The quark mass matrices with the form (1)
are diagonalized by the bi-unitary transformation
$D_f  =  U_{Lf}^\dagger M_f U_{Rf}$, 
where $U_{Lf}\equiv P_{Lf}^\dagger O_f$, 
$U_{Rf}\equiv P_{Rf}^\dagger O_f$, 
and $O_d$ ($O_u$) is given by 
\begin{equation}
O_f \equiv
\left(
\begin{array}{ccc}
{ c_f}& { s_f}& {0} \\
{-\frac{s_f}{\sqrt{2}}}&
{\frac{c_f}{\sqrt{2}}}&
{-\frac{1}{\sqrt{2}}} \\
{-\frac{s_f}{\sqrt{2}}}&
{\frac{c_f}{\sqrt{2}}}&
{\frac{1}{\sqrt{2}}}
\end{array}
\right) \ . \
%(5)
\end{equation}   
(For simplicity, hereafter, we will take $P_R=P_L^\dagger$,
so that the matrix $M_f$ becomes a symmetric matrix.
However, this assumption is not essential for the results
in the present model.) 
Then, the Cabibbo--Kobayashi--Maskawa (CKM)  quark mixing 
matrix $V$ is given by
\begin{eqnarray}
V&=&U^\dagger_{Lu}U_{Ld}=O^T_uP_{u}P^\dagger_{d} O_d\nonumber\\[.1in]
& =&
\left(
\begin{array}{ccc}
c_uc_d+\rho s_u s_d \quad & c_u s_d-\rho s_u c_d \quad & -{\sigma}s_u \\
s_uc_d-\rho c_u s_d \quad & s_u s_d+\rho c_u c_d \quad & {\sigma}c_u \\
-{\sigma}s_d \quad & {\sigma}c_d \quad & \rho \\
\end{array}
\right),\label{eq-ourckm} 
%(6)
\end{eqnarray}
where  \(\rho\) and \(\sigma\) are defined by 
\begin{equation}
\rho=\frac{1}{2}(e^{i\delta_3}+e^{i\delta_2}) 
\ , \ \ \ 
%=\cos\frac{\delta_3 - \delta_2}{2} \exp i
%\left( \frac{\delta_3 + \delta_2}{2} \right) \ ,
%\end{equation}
%\begin{equation}
\sigma=\frac{1}{2}(e^{i\delta_3}-e^{i\delta_2})
%= \sin\frac{\delta_3 - \delta_2}{2} 
%\exp i \left( \frac{\delta_3 + \delta_2}{2}+ \frac{\pi}{2}
%\right) 
\ . 
%(7)
\end{equation}
Here we have put \(P \equiv P_{u}P^\dagger_{d} \equiv 
\mbox{diag}(e^{i\delta_1}, e^{i\delta_2},e^{i\delta_3})\), and
we have taken \(\delta_1=0\) without 
loss of generality.

{}From the expression (6),
we obtain the phase-parameter independent predictions
(the 3rd generation quark-mass independent
predictions \cite{Branco})
\begin{equation}
\frac{\left|V_{ub}\right|}{\left|V_{cb}\right|} =\frac{s_u}{c_u} 
=\sqrt{\frac{m_u}{m_c}} 
%=\sqrt{\frac{2.33}{677}}=0.0586\pm 0.0064 
\ , \ \ 
\frac{\left|V_{td}\right|}{\left|V_{ts}\right|} =\frac{s_d}{c_d}
=\sqrt{\frac{m_d}{m_s}} \ ,
%=\sqrt{\frac{4.69}{93.4}}=0.224\pm 0.014 \ ,
%\eqno(8)
\end{equation}
which are almost independent of the RGE effects. 

Next let us fix the parameters $\delta_3$ and $\delta_2$.
{}From the relation 
\begin{equation}
|V_{cb}|=  \frac{1}{ \sqrt{1+{m_u}/{m_c}} }
\sin\frac{\delta_3-\delta_2}{2} \ ,
%\eqno(9)
\end{equation}
and the observed value \cite{PDG} $|V_{cb}|=0.0412\pm 0.0020$,
we obtain $\delta_3 -\delta_2 = 4.59^\circ\pm 0.21^\circ$.
Also, from the expression of $|V_{us}|$,
we can obtain the value $\delta_3+ \delta_2 = 
93^\circ \pm 22^\circ$. 
Because of the small value 
$\sin(\delta_3-\delta_2)/2 \simeq 0.04$,  
we obtain the following approximate relations
\begin{equation}
\left| V_{us}\right|
% =c_u s_d\left|1 -\rho{\frac{s_u}{c_u}}{\frac{c_d}{s_d}}\right|
\simeq\sqrt{\frac{m_d}{m_s}} \ ,
\ \ \ 
\left| V_{cd}\right|
% = c_u s_d\left|\rho -{\frac{s_u}{c_u}}{\frac{c_d}{s_d}}\right|
\simeq\sqrt{\frac{m_d}{m_s}} \ ,
\  \ \ 
\left| V_{td}\right|  %= \left| {\sigma} \right| s_d
\simeq \left|V_{cb}\right|{\cdot} \left|V_{us}\right| \ ,
%\eqno(10)
\end{equation}
which are consistent with the present experimental data
\cite{PDG}.

%%%%%%%%%%%%%%%%%%%%%%%
In the present model, 
the rephasing invariant Jarlskog 
parameter $J$ is given by
\begin{equation}
J   =  |\sigma|^2|\rho|c_u s_u c_d s_d \sin \frac{\delta_3+\delta_2}{2} 
 \simeq  |V_{ub}||V_{cb}||V_{us}|\sin\frac{\delta_3+\delta_2}{2}.
%\eqno(11)
\end{equation}
Therefore, the phase factor $\delta$  in the standard expression 
of $V$  corresponds to $\delta \simeq {\delta_3+\delta_2}/{2}$
in the present model.
We predict $|J|= (1.91 \pm 0.38) \times 10^{-5}$.

%%%%%%%%%%  chap 4  %%%%%%%%%%%%%%
\section{Lepton mixing matrix}
We assume that the neutrino masses are generated via 
the seesaw mechanism $M_\nu=-M_D M_R^{-1} M_D^{T}$.
Here $M_D$ and $M_R$ are the Dirac neutrino and 
the right-handed Majorana neutrino mass matrices.
Note that when we assume the same matrix forms (1)
for $M_D$ and $M_R$,
the effective neutrino mass matrix $M_\nu=-M_D M_R^{-1}M_D$
is again given by the same texture (1): 
\begin{equation}
M_\nu  = -P_\nu^\dagger \widehat{M}_D \widehat{M}_R^{-1}
\widehat{M}_D^T P_\nu^\dagger 
= P_\nu^\dagger \widehat{M}_\nu P_\nu^\dagger \ .
%\eqno(12)
\end{equation}
Therefore, we obtain the lepton mixing matrix $U$
\begin{equation}
U  =O_e^T P O_\nu 
 =
\left(
\begin{array}{ccc}
c_ec_\nu+ \rho_\nu s_e s_\nu \quad 
& c_e s_\nu- \rho_\nu s_e c_\nu 
\quad & - \sigma_\nu s_e \\
s_ec_\nu- \rho_\nu c_e s_\nu \quad 
& s_e s_\nu+ \rho_\nu c_e c_\nu 
\quad & \sigma_\nu c_e \\
- \sigma_\nu s_\nu \quad & \sigma_\nu c_\nu \quad 
& \rho_\nu \\
\end{array}
\right), 
%\eqno(13)
\end{equation}
where $P \equiv P_{e} P_\nu^\dagger \equiv 
\mbox{diag}(e^{i\delta_{\nu1}},
e^{i\delta_{\nu2}},e^{i\delta_{\nu3}})$.
Hereafter we will again take 
$\delta_{\nu1}=0$ without loss of generality.
Note that
$V=O_u^T P O_d$,
while
$U=O_e^T P O_\nu$,
so that 
all the mixing formulae in the lepton sectors
are given by the replacement $(m_u, m_c, m_t) \rightarrow
(m_e, m_\mu, m_\tau)$ and $(m_d, m_s, m_b) \rightarrow
(m_{\nu 1}, m_{\nu 2}, m_{\nu 3})$ 
in those in the quark sectors.
However, this does not means that the physics in the up-quark
(down-quark) sector corresponds to the physics in the charged
lepton (neutrino) sector.
In fact, as we see in Table 1, the parameter values in
each sector  are different from the other.

%%%%%%%%%%%%%%%%%%%%%%
%%%%%%%%%%%%%%%%%%%%%%%

We obtain the phase-parameter independent predictions
\begin{equation}
\frac{\left|U_{13}\right|}{\left|U_{23}\right|} = \frac{s_e}{c_e}
=\sqrt{\frac{m_e}{m_\mu}} 
%=\sqrt{\frac{0.487}{103}}=0.0688, 
\ , \ \ \ 
\frac{\left|U_{31}\right|}{\left|U_{32}\right|} = \frac{s_\nu}{c_\nu}
=\sqrt{\frac{m_{\nu 1}}{m_{\nu2}}} \ .
%\eqno(14)
\end{equation}
The neutrino mixing angle 
$\theta_{atm}$ 
under the constraint
$ |\Delta m^2_{23}|\gg |\Delta m^2_{12}|$  is given by
\begin{equation}
\sin^2 2 \theta_{\mbox{{\tiny atm}}} 
 \equiv 
4 \left| U_{23} \right|^2 \left|U_{33} \right|^2 =
4\left|\rho_\nu \right|^2 \left|\sigma_\nu\right|^2 c^2_e 
= \frac{m_{\mu}}{m_{\mu}+m_e}\sin^2(\delta_{\nu3}-\delta_{\nu2})
 \ .
%\eqno(15)
\end{equation}
We assume the maximal mixing between $\nu_\mu$ and $\nu_\tau$, 
so that we take $\delta_{\nu3}-\delta_{\nu2}={\pi}/{2}$.
Then, the model predicts 
\begin{equation}
|U_{13}|^2={\frac{1}{2}}{\frac{m_e}{m_\mu+m_e}}=0.00236  \ ,
%\eqno(16)
\end{equation}
which is consistent with the  constraint
$|U_{13}|_{\mbox{\tiny exp}}^2 <  0.03$ from the CHOOZ data
\cite{chooz}.
The mixing angle $\theta_{solar}$  
is given by
\begin{equation}
\sin^2 2\theta_{solar}
\equiv 
4 \left|U_{11}\right|^2 \left|U_{12}\right|^2 
\simeq  \frac{4m_{\nu 1}/m_{\nu 2}}{(1+m_{\nu 1}/m_{\nu 2})^2} \ , 
%\eqno(17)
\end{equation}
which leads to the relation 
$m_{\nu 1}/m_{\nu 2} \simeq \tan^2\theta_{solar}$.
Therefore, the best fit value \cite{sno} $\tan^2\theta_{solar}=0.34$
predicts the neutrino mass ratio ${m_{\nu 1}}/{m_{\nu 2}} 
\simeq 0.34$, so that we can obtain the neutrino masses
\begin{equation}
m_{\nu 1}  =  0.0030 \, {\rm eV} \ ,\ \
m_{\nu 2}  =  0.0088 \, {\rm eV} \ ,\ \ 
m_{\nu 3}  =  0.050 \, {\rm eV} \ , 
%\eqno(18)
\end{equation}
where we have used the best fit values \cite{kamland,skamioka} of
$\Delta m^2_{solar}=6.9 \times 10^{-5}$ eV$^2$
and $\Delta m^2_{atm}=2.5\times 10^{-3}$ eV$^2$.

We also obtain the averaged neutrino mass 
$\langle m_\nu \rangle \sim (10^{-3} - 10^{-4})$ eV,
but the explicit value is highly dependent on 
the value of
$\delta_\nu\equiv (\delta_{\nu 3} +\delta_{\nu 2})/2$.
At present, we cannot fix the value of $\delta_\nu$.

%%%%%%%%%%%%%%%%%%%%%%%%%%%%%%%%%%%%%%%%%%

\section{Conclusion}

In conclusion, stimulated by recent neutrino data, which suggest
a nearly bimaximal mixing, 
we have investigated a possibility that 
all the mass matrices of quarks and leptons have the same
texture as the neutrino mass matrix.  
In spite of the assumption of the
universal texture for all the fermion mass matrices, 
we can obtain the differences between $V_{quark}$ and
$V_{lepton}$ as follows:
(i) the mixing between 1st and 2nd generations is
given by $\tan\theta_{12} = \sqrt{{m_1}/{m_2}}$, so that
the well-known empirical relation $|V_{us}|\simeq \sqrt{m_d/m_s}$ 
is due to the observed mass rations $m_u/m_c \ll m_d/m_s \ll 1$, 
and the nearly maximal mixing 
$|V_{e2}| \sim 1/\sqrt{2}$
is due to the approximate degeneracy $m_{\nu 1} \sim m_{\nu 2}$
and the observed mass ratio $m_e/m_\mu \ll 1$; 
(ii) the mixing between 2nd and 3rd generations is given by
the relation (9) (and the corresponding relation in the
lepton sector), so that the small value $|V_{cb}|\simeq 0.04$
means $(\delta_3-\delta_2)/2 \simeq 0.04$ and $m_u/m_c \ll 1$,
and the maximal mixing $V_{\mu 3}\simeq 1/\sqrt{2}$ means
$\delta_3-\delta_2 \simeq \pi/2$ together with $m_e/m_\mu
\ll 1$.

The present data in the quark sectors have already
fixed the $CP$ violating phase parameters
$\delta_3$ and $\delta_2$, while
the present neutrino data have yet not fixed
the parameter 
$(\delta_{\nu 3} +\delta_{\nu 2})/2$,
although they have fixed the value of 
$(\delta_{\nu 3} -\delta_{\nu 2})/2$.
We hope that  future experiments on the
$CP$ violation will fix our remaining parameter
$\delta_\nu$.
Then, we will be able to obtain a clue to the
origin of our phase parameters $\delta_i$
($\delta_{\nu i}$).

%%%%%%%%%%%%%%%%%%%%%%%%%%%%%%%%%%%%%
Since, in the present model, each mass matrix
$M_f$ (i.e. the Yukawa coupling $Y_f$) takes
different values of the parameters $a_f$, $b_f$, 
and so on,
the present model cannot be embedded into a GUT
scenario.
In spite of such a demerit, however,
it is worth while noting that 
the present model can give 
a unified description of quark and
lepton mass matrices with the same texture.

%%%%%%%%%%%%%%%%%%%%%%% ref %%%%%%%%%%%%%%%%%%%%%%%%%%%%%%%%%%%%%%%%%%%

%%%%%%%%%%%%%%%%%%%%%%%%%%%%%%%%%%%%%%%%%%%%%%%%%
%%%%%%%%%%%%%%%%%%%%%%%%%%%%%%%%%%%%%%%%%%%%%
\begin{table}[h]
%\caption{
\begin{quotation}
Table 1. \ 
Input values $m_i$ and output values $a_f$, $b_f$ and $c_f$.
The numerical values are given in unit of GeV except for 
the neutrino sector (in unit of eV).
The input values are those \cite{F-K} at $\mu=m_Z$ except for
the neutrino sector (for the neutrino sector, see the text).
\vspace*{1pt}
%}
\end{quotation}

%{\footnotesize
%\tabcolsep7pt
%\begin{tabular}{@{}crrrrrr@{}}
\begin{tabular}{|c|ccc|ccc|}
\hline
{} &{} &{} &{} &{} &{} &{}\\[-1.5ex]
{} &{} &{Inputs} &{} &{} &{Outputs} &{}\\[1ex]
{$f$} & $m_{f1}$ & $m_{f2}$ & $m_{f3}$ & $a_f$ & $b_f$ & $-c_f$ \\[1ex]
\hline
{} &{} &{} &{} &{} &{} &{}\\[-1.5ex]
$u$ & 0.00233 & 0.677 & 181 & 0.0280 & 90.8 & 90.2 \\[1ex]
$d$ & 0.00469 & 0.0934 & 3.00 & 0.0148 & 1.54 & 1.46 \\[1ex]
$e$ & 0.000487 & 0.103 & 1.75 & 0.00500 & 0.924 & 0.822 \\[1ex]
$\nu$ & 0.0030 & 0.0088 & 0.050 & 0.0036 & 0.028 & 0.022 \\[1ex] 
\hline
\end{tabular}\label{tab1} 
%}
\vspace*{-13pt}
\end{table}
%%%%%%%%%%%%%%%%%

\end{document}